\documentclass[epj]{svjour}
\usepackage{graphics}
\begin{document}
\title{Development of non-modal shear induced instabilities in atmospheric tornadoes}
\author{M. Abuladze \and Z. Osmanov
}                     
\offprints{}          
\institute{School of Physics, Free University of Tbilisi, 0183, Tbilisi,
Georgia}
\date{Received: date / Revised version: date}
%
\abstract{
In this paper we consider the role of nonmodal instabilities in the dynamics of atmospheric tornadoes. For this purpose we consider the Euler equation, continuity equation and the equation of state and linearise them. As an example we study several different velocity profiles: the so-called Rankine vortex model; the Burgers-Rott vortex model; Sullivan and modified Sullivan vortex models. It has been shown that in the two dimensional Rankine vortex model no instability appears in the inner region of a tornado. On the contrary, outside this area the physical system undergoes strong exponential instability. We have found that initially perturbed velocity components lead to amplified sound wave excitations. The similar results have been shown in Burgers-Rott vortex model as well. As it was numerically estimated, in this case, the unstable wave increases its energy by a factor of $400$ only in $\sim 0.5$min. According to the numerical study, in Sullivan and modified Sullivan models, the instability does not differ much by the growth. Despite the fact that in the inner area the exponential instability does not appear in a purely two dimensional case, we have found that in the modified Sullivan vortex even a small contribution from vertical velocities can drive unstable nonmodal waves. 
\PACS{
      {PACS-key}{discribing text of that key}   \and
      {PACS-key}{discribing text of that key}
     } 
} 
\maketitle
\section{Introduction}
\label{Theory}


It is evident, that many astrophysical, geophysical or laboratory flows reveal complex velocity fields, the so-called shear flows (SF). In the context of the present paper a special interest deserve atmospheric tornadoes. For measuring the velocity profile the so-called Doppler radars are normally used \cite{doppler1,doppler2} and according to the observations atmospheric tornadoes reveal very complex velocity configuration, which in turn, means that they are characterised by velocity SFs. Therefore, it is interesting to study the possible influence of these kinematic complexities on the overall dynamics of tornadoes.

In general, it is natural that in velocity SFs strongly unstable waves might develop \cite{modal}. In the framework of standard, modal approach, the linear analysis is performed by implying an anzatz, where the time dependence of excited unstable waves are fixed. Unlike the mentioned class of waves, the non-modal instabilities might also be driven by means of the velocity shear. This particular class of modes have been studied and corresponding mathematical model has been developed by Lord Kelvin \cite{kelvin}. In this approach, the linearly unstable waves are examined and the system of equations governing the overall dynamics of waves is reduced to the initial value problem with ordinary differential equations \cite{tref}.

As it has been realised, even very simple flows reveal quite non trivial behaviour. In particular in \cite{couette} the authors have considered a compressible Cuette flow and linearising the hyrodynamyc equations performed the nonmodal analysis. It has been shown that sound-type perturbations exchange energy with the mean flow. Again, very simple geometry of unperturbed velocity field has been examined in \cite{vazha}, where the linear perturbation dynamics was studied for non relativistic electron-positron plasmas and it has been found that under certain circumstances velocity shear might drive the Alfv\'en waves. Generally speaking a series of articles has been dedicated to the study of nonmodal waves in magnetohydrodynamic (MHD) plasmas. In particular, in \cite{chven} It has been found that in MHD helical flows the  Alfv\'en waves and slow/fast magnetosonic waves blend with each other and become nonmodally unstable. Another interesting phenomenon has been discovered. In the framework of the theoretical model, one can show that depending on shear parameters the wave vector components might evolve in time either exponentially or harmonically. Consequently, in the former case the physical system leads to excitation of nonmodally unstable waves with relatively high growth rates. On the other hand, when the dependance on time is harmonic, the generated modes are normally stable, but for certain, usually narrow range of parameters, the excited waves become unstable. The similar study has been performed for electrostatic waves \cite{electrost}, where apart from considering the exponentially and parametrically unstable waves it has been shown that under certain circumstances the physical system is subject to the excitation of beat wave phenomena. The possible application of the nonmodal instabilities in the so-called self heating mechanism was examined in a series of articles \cite{andro,rop10,orp12}.

In the above cited papers the authors mainly studied the nonmodal waves in a purely astrophysical context. But we have already discussed, that the atmospheric tornadoes are strongly influenced by the velocity SFs and therefore the scope of the resent paper is to study the role of the mentioned class of hydrodynamic waves in the mean flow of atmospheric tornadoes.

The paper is organized in the following way. In section~2, a theoretical model of shear-induced instability is presented, in section~3 considering several vortex-like velocity structures we discuss the obtained numerical results and in section ~4 we summarise them.

\section{Main consideration}

In this section we present the mathematical model of the nonmodal analysis. For this purpose we write the equations that govern the mentioned physical system. These are the continuity equation
\begin{equation}
\label{conn} D_t\rho + \rho \nabla \cdot {\bf  V}= 0,
\end{equation}
the Euler equation:
\begin{equation}
\label{euler} D_t{\bf V} = -\frac{1}{\rho}{\bf \nabla }P +{\bf g},
\end{equation}
and the polytropic equation of state:
\begin{equation}
\label{eqstate} P=C{\rho}^{n},
\end{equation}
where $D_t \equiv \partial_t + ({\bf V} \cdot \nabla)$,
$\rho$ is the density, ${\bf V}$
is the velocity, $P$ denotes pressure,  ${\bf g}$ is the free fall acceleration and $n$ is the polytropic index.

For studying the development of nonmodal waves we expand all physical variables around their equilibrium values, preserving only first order terms:

\begin{equation}
\label{psi} \Psi \approx \Psi_0 + \Psi',
\end{equation}
where $\Psi \equiv \{\rho, {\bf V}, P\}$ and throughout the paper we assume that the unperturbed density is constant.

By combining Eqs. (\ref{conn}-\ref{psi}) one obtains 
\begin{equation}\label{con}
\mathcal{D}_t\rho' + \rho_0 (\nabla \cdot {\bf V'})= 0,
\end{equation}
\begin{equation}\label{eul}
\mathcal{D}_t{\bf V'} + ({\bf V'} \cdot \nabla){\bf V_0} =
-\frac{C_s}{\rho_0}{\bf \nabla \rho'},
\end{equation}
where $\mathcal{D}_t \equiv \partial_t + ({\bf V_0}\cdot \nabla)$
and $C_s = \sqrt{dP_0/d\rho_0}$ is the sound speed.

The development of nonmodal SF instabilities is a direct result of velocity inhomogeneity. We follow the standard method and expand the velocity in a Taylor series close to any point $A(x_0, y_0, z_0)$ preserving first order terms
\begin{equation}\label{velexpand}
{\bf V}={\bf V}(A)+\sum_{i=1}^3\frac{\partial{\bf V}(A)}{\partial
x_i}(x_i-x_{i0}),\end{equation}
where $i=1,2,3$ and $x_i=(x,y,z)$. Then by implying the anzatz
\begin{equation}\label{anzatz}
F(x,y,z,t)\equiv\hat{F}(t)e^{\phi_1-\phi_2},\end{equation}
\begin{equation}\label{fi1}
\phi_1\equiv\sum_{i=1}^3{K_i}(t)x_i,\end{equation}
\begin{equation}\label{fi2}
\phi_2\equiv\sum_{i=1}^3V_i(A)\int{K_i}(t)dt,\end{equation}
where by $V_i(A)$ we denote the unperturbed velocity components and $K_i(t)$'s
are the wave vector components, which if satisfy the equations:
\begin{equation}\label{dk}
{\bf \partial_{t}K}+ {\bf S^T} \cdot {\bf K}=0,\end{equation}
with  ${\bf S}$ :
\begin{equation}\label{S}
 {\bf S} = \left(\begin{array}{ccc} V_{x,x} & V_{x,y} & V_{x,z}  \\
V_{y,x} & V_{y,y} & V_{y,z}  \\ V_{z,x} & V_{z,y} & V_{z,z} \\
\end{array} \right ),\end{equation}
where $V_{i,k}\equiv\partial V_{i}/\partial x_k$ and ${\bf S^T}$ is the transposed matrix, lead the original equations for the perturbed quantities to the following dimensionless set of equations (we follow the model developed in \cite{rop10,orp12})
\begin{equation}\label{cont}
d^{(1)}+{\bf k} \cdot {\bf u} = 0,\end{equation}
\begin{equation}\label{v}
{\bf u}^{(1)}+{\bf s}\cdot{\bf u} = {\bf k} d,\end{equation}
\begin{equation}\label{k}
\textbf{k}^{(1)} + {\bf s^T} \cdot {\bf k} = 0, \end{equation}
where we have used the notations: $\textbf{u} \equiv \textbf{V}'/C_s$, $d \equiv
-i\rho'/\rho_0$, $\textbf{k}\equiv \textbf{K}/K_x(0)$, $\textbf{s}\equiv \textbf{S}/(K_x(0)C_s)$.

For the study of the development of instabilities it is natural to examine energy of the excited waves, composed of the so-called kinetic and compressional terms \cite{landau}
\begin{equation}\label{E}
E_{tot}=E_{kin}+E_c=\frac{\textbf{u}^2}{2}+\frac{d^2}{2}.
\end{equation}

\section{Discusion}

In this section we examine several models of tornadoes studying the generation of nonmodal waves. For this purpose we consider the realistic form of the helical flow written in cylindrical coordinates
\begin{equation}\label{flow}
\textbf{V} = \{0, \upsilon(r), U(r)\},        
\end{equation}
where $r$ is the radial coordinate.
The aforementioned choice reduces the shear matrix to the following expression
\begin{equation}\label{S}
 {\bf S} = \left(\begin{array}{ccc} 0 & -\omega & \;\;\;\;0  \\
 \\
r\frac{\partial\omega}{\partial r}+\omega & \;0 & \;\;\;\;0  \\ 
\\
\frac{\partial U}{\partial r} & \;0 & \;\;\;\;0 \\
\end{array} \right ),\end{equation}
where $\omega\equiv\upsilon/r$ is the angular velocity of rotation.
From Eqs. (\ref{cont}-\ref{k}) it is evident that the overall dynamics of perturbations strongly depends on shear matrix. Therefore, it is interesting to examine several vortex configurations modelling flows in tornadoes.

\subsection{Rakine vortex}

Kinematic parameters of tornadoes differ from one particular tornado to another. Without going into details, we consider typical kinematic parameters for tornadoes. In particular, from observations it is known that the maximum tangential velocity of rotation might be of the order of $8\times 10^3$cm s$^{-1}$ and the radius, $R$, of the tornadoe - of the order of $5\times 10^4$cm \cite{parameters}. 

As a first example we consider the simplest velocity configuration the so-called Rankine vortex. In the framework  of this model tangential velocity  linearly increases from zero to a certain maximum radius, $r_m$, (characterising a rigidly rotating region of a tornado) and then decreases as inversely proportional to the radial distance \cite{rankine}
\begin{equation}\label{rankine}
\upsilon(r)=\upsilon_m\left(\frac{r}{r_m}\right)^{\alpha},
\end{equation}
where $\alpha=1$ for $r\leq r_m$ and $\alpha=-1$ for $r > r_m$. In the first approximation there is no vertical component of velocity, $U(r)=0$, which in turn reduces the problem to two-dimensional case with the matrix
\begin{equation}\label{S1}
 {\bf S} = \left(\begin{array}{ccc} 0 & -\omega \\
 \\
r\frac{\partial\omega}{\partial r}+\omega & \;0  \\ 

\end{array} \right ),\end{equation}

\begin{figure}
\resizebox{0.5\textwidth}{!}{%
  \includegraphics{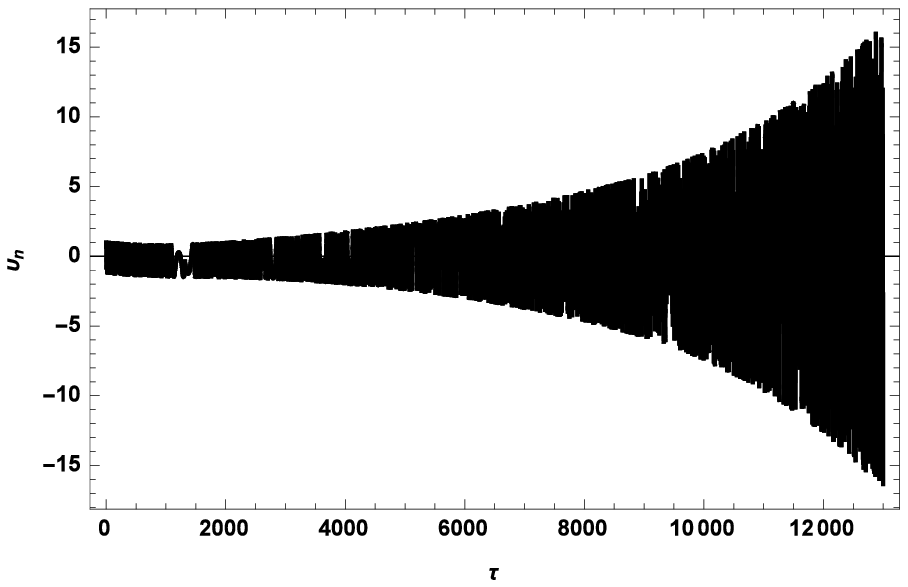}
  \includegraphics{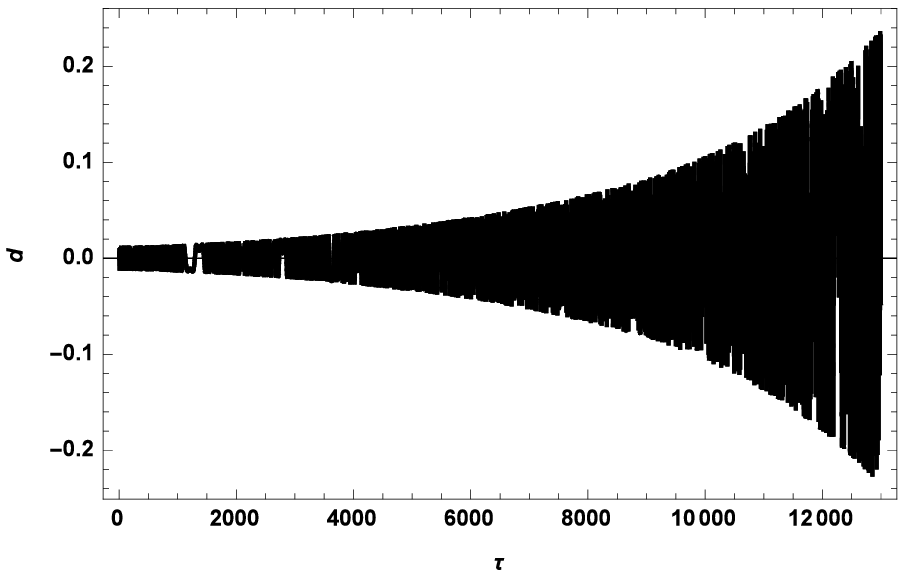}

}
\caption{In the outer region, $r = R/2$, of a tornado for the Rankine model the temporal evolution of normalised total velocity perturbation (left panel) and density perturbation (right panel) are shown. The set of parameters is $r_m=R/3$, $\upsilon_m = 8\times 10^3$cm s$^{-1}$, $\lambda_0=r_m/100$, $\upsilon_{x,y}(0) = 0.01$, $d(0) = 0$, $k_x(0) = 1$ and $k_y(0) = 0.1$. By $\lambda_0$ we denote the initial wavelength of the perturbation.}
\label{fig1}       
\end{figure}

From Eq. (\ref{k}) in the two dimensional case (without the $z$ component of velocity) one can straightforwardly derive the following equations for the components of the wave vector 
\begin{equation}\label{k_1}
k^{(2)}_{x,y}-\Gamma^2k_{x,y}=0,
\end{equation}
where $\Gamma^2\equiv -\omega\left(\omega+r\frac{\partial\omega}{\partial r}\right)$. It is clear that in the "inner" region, $r\leq r_m$,  $\Gamma^2$ is negative and therefore the wave vector components behave harmonically and the system does not undergo relatively fast exponential instabilities. In general, there is a possibility of the so-called parametric instability despite the mentioned condition \cite{orp12}, but usually its increment is not high and we do not consider it in the present paper. Unlike this case, in a region $r > r_m$,  $\Gamma^2$ becomes positive and consequently the wave components undergo exponential growth, inevitably leading to the unstable behaviour of excited hydrodynamic waves.

In particular, in Fig. \ref{fig1} we show the time dependence of total velocity perturbation, $\upsilon_n\equiv {\bf \mid u\mid}/{\bf \mid u_0\mid}$ (left panel) normalised on its initial value and density perturbation, $d$, (right panel) in the outer region $r=R/2$. The set of parameters is $r_m=5\times 10^4$cm, $\upsilon_m = 8\times 10^3$cm s$^{-1}$, $\lambda_0=r_m/100$, $\upsilon_{x,y}(0) = 0.01$, $d(0) = 0$, $k_x(0) = 1$ and $k_y(0) = 0.1$, where $\lambda_0$ is the initial wavelength of the perturbation. As it is clear from the parameters initially only velocity components are perturbed. As a result the physical system leads to the density perturbation (right panel) and therefore, sound waves are generated. We see from the plots that the excited nonmodal velocity shear induced waves exhibit efficient instability.

\begin{figure}
\resizebox{0.5\textwidth}{!}{%
  \includegraphics{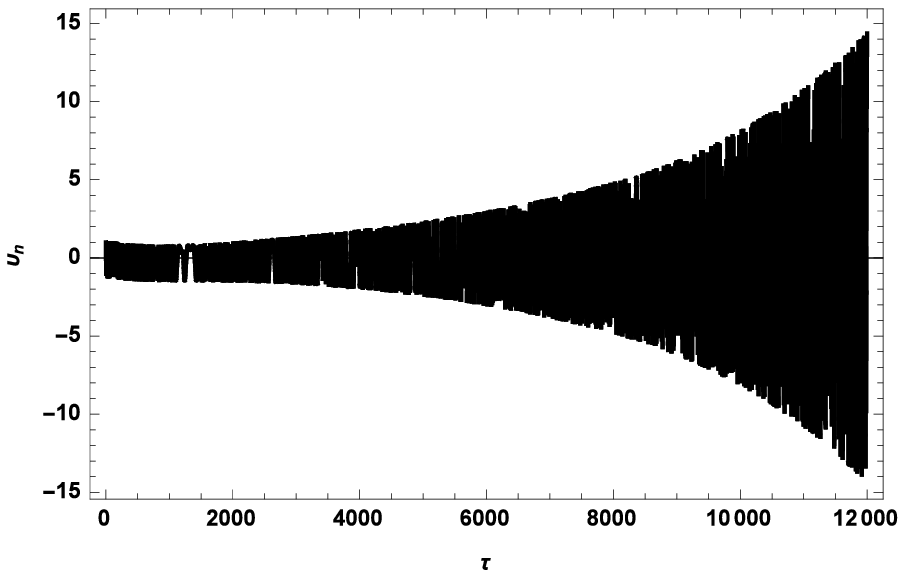}
  \includegraphics{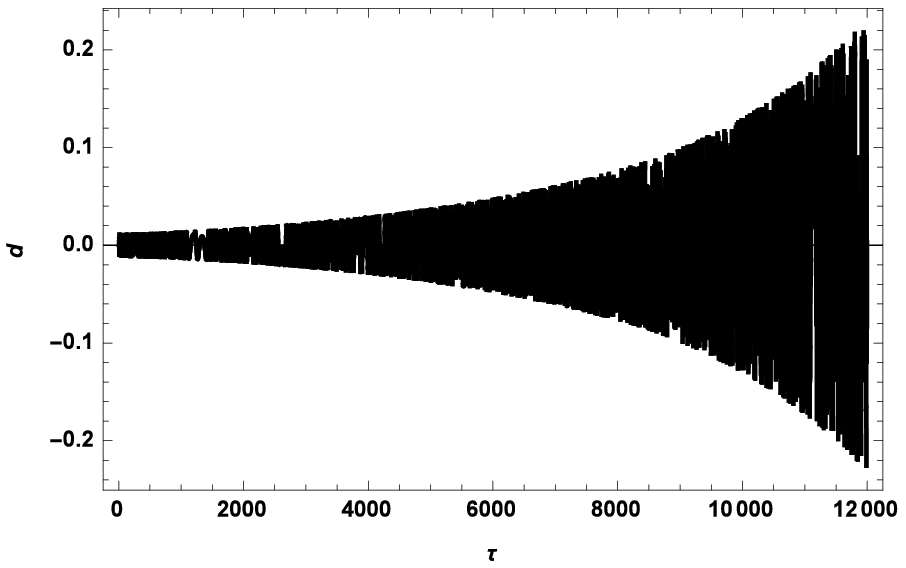}

}
\caption{In the outer region of a tornado for the Burgers-Rott vortex the temporal evolution of a normalised value of total velocity perturbation (left panel) and density perturbation (right panel) are shown. The set of parameters is the same as in the previous figure. }
\label{fig2}       
\end{figure}
\subsection{Burgers-Rott vortex}

According to the recent Doppler observations it becomes evident that the so-called one-celled atmospheric tornadoes can be described better by the Burgers-Rott vortex than by the Rankine model \cite{rankine}
\begin{equation}\label{burg}
\upsilon(r)=1.4\upsilon_m\left(\frac{r}{r_m}\right)^{-1}\left[1-exp\left\{-1.2564\left(\frac{r}{r_m}\right)^2\right\}\right].
\end{equation}
In this model the velocity configuration does not allow the appearance of an unstable behaviour of the wave vector in the inner zone, $r\leq r_m$ and like the previous case outside this area the modes are efficiently amplified.

In Fig. \ref{fig2} we present the temporal dependance of a normalised value of total velocity perturbation, $\upsilon_n$ (left panel) and density perturbation, $d$, (right panel) in the outer region $r=R/2$. The set of parameters is the same as in the Fig. \ref{fig2}. As it is clear, from the plots, in the framework of the Burgers-Rott model the general behaviour is similar to the previous case. In particular, initially excited velocity components lead to the sound wave excitation. 

\begin{figure}
\resizebox{0.5\textwidth}{!}{%
  \includegraphics{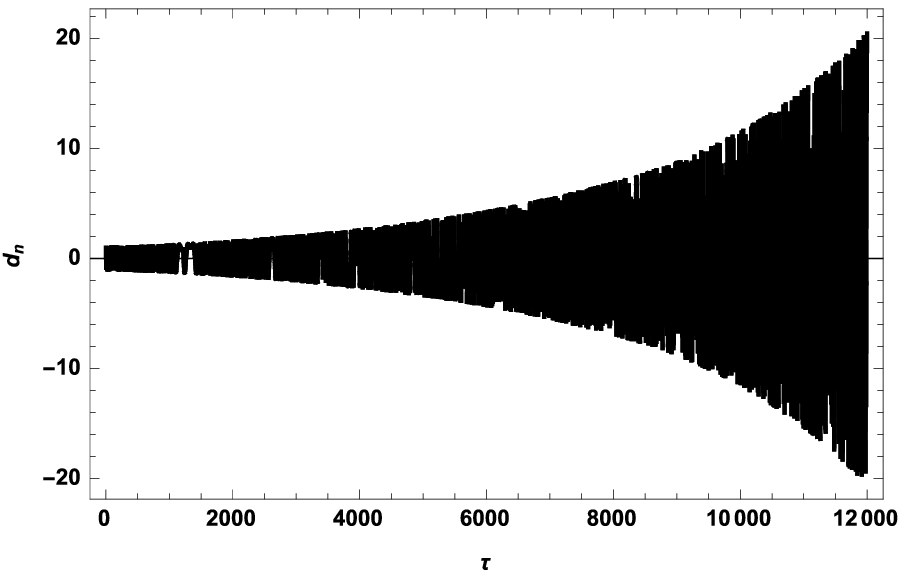}
   \includegraphics{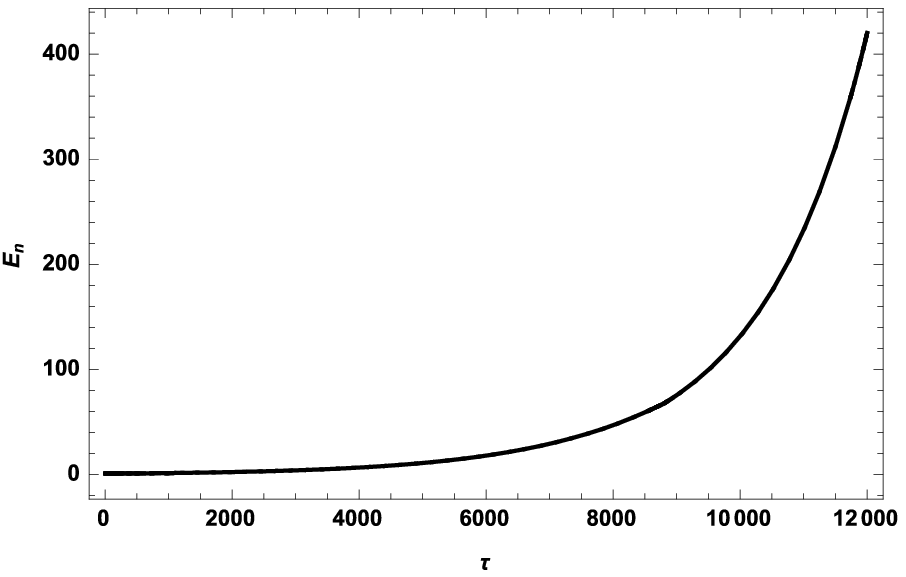}

}
\caption{In the outer region, $r = R/2$, of a tornado for the Burgers-Rott vortex the temporal  evolution of normalised values of density perturbation (left panel) and energy perturbation (right panel) are shown. The set of parameters is $r_m=R/3$, $\upsilon_m = 8\times 10^3$cm s$^{-1}$, $\lambda_0=r_m/100$, $\upsilon_{x,y}(0) = 0$, $d(0) = 0.01$, $k_x(0) = 1$ and $k_y(0) = 0.1$. }
\label{fig3}       
\end{figure}

Amplification of physical quantities, $\upsilon_{x,y}$ and $d$,  means that the generated wave increases its energy. In Fig. \ref{fig3} we show this particular dependence, but unlike the previous cases, here initially the sound waves are excited. The set of parameters is the same as in the previous cases except $\upsilon_{x,y}(0) = 0$, $d(0) = 0.01$. As we see, for the aforementioned physical parameters energy of the non-modal waves increases by the factor of two orders of magnitude with respect to its initial value. One also can straightforwardly check that the generated sound modes lead to the perturbation of velocity components.

\subsection{Sullivan and modified Sullivan vortex}
\begin{figure}
\resizebox{0.5\textwidth}{!}{%
  \includegraphics{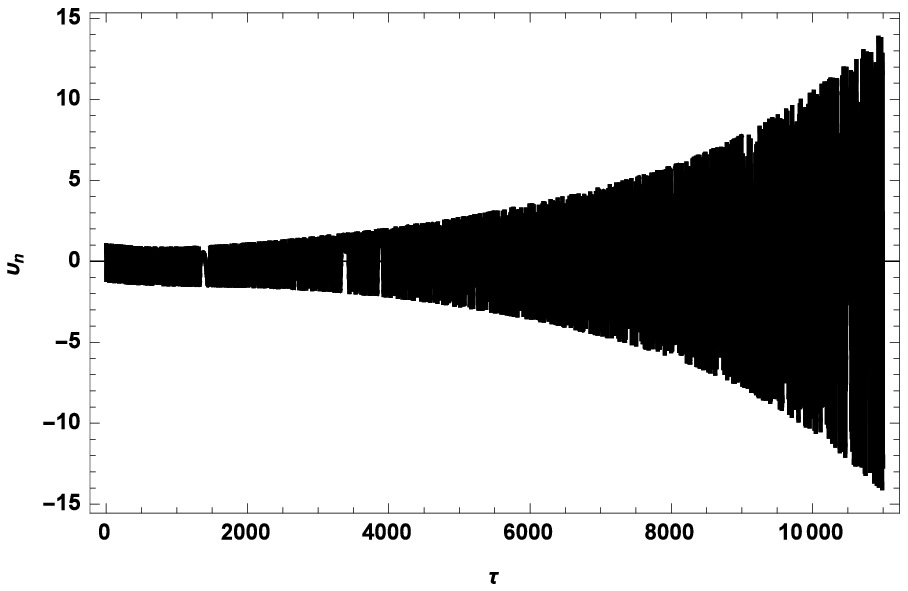}
  \includegraphics{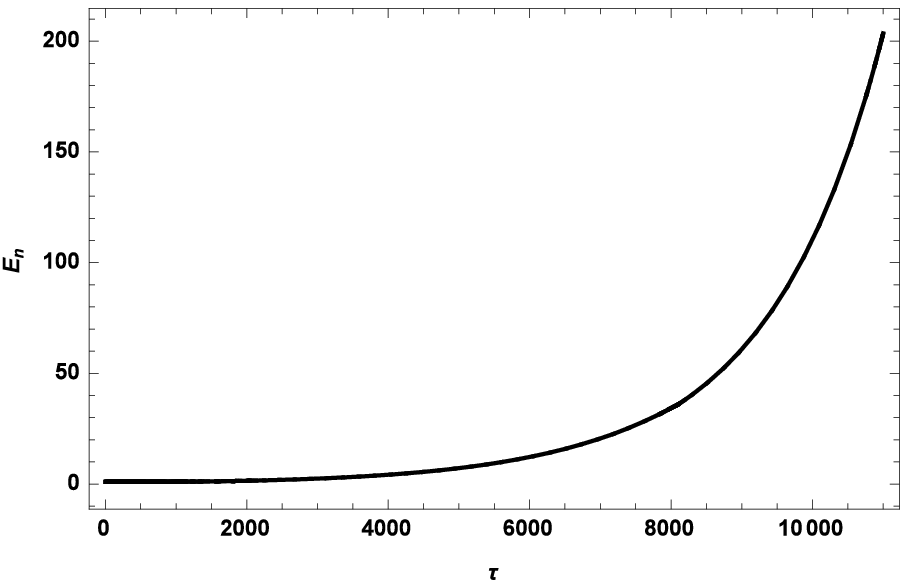}

}
\caption{In the outer region, $r = R/2$, of a tornado for the Sullivan vortex the temporal evolution of normalised values of total velocity perturbation (left panel) and energy perturbation (right panel) are shown. The set of parameters is $\alpha=2.4$, $\beta=7.89$, $\gamma=-0.435$, $A = 0.3$ and $B=0.7$, $r_m=5\times 10^4$cm, $\upsilon_m = 8\times 10^3$cm s$^{-1}$, $\lambda_0=r_m/100$, $\upsilon_{x,y}(0) = 0.01$, $d(0) = 0$, $k_x(0) = 1$ and $k_y(0) = 0.1$.}
\label{fig4}
\end{figure}

Measurements by using the Doppler radars strongly indicate that two-celled tornadoes can be characterised by the following profile of the tangential velocity (Sullivan vortex)
\begin{equation}\label{sullivan}
\upsilon(r)=\upsilon_m\left(\frac{r}{r_m}\right)^{\alpha}\left[A+B\left(\frac{r}{r_m}\right)^{\beta}\right]^{\gamma},
\end{equation}

where $\alpha=2.4$, $\beta=7.89$, $\gamma=-0.435$, $A = 0.3$ and $B=0.7$. In the framework of this model in the inner region, $r\leq r_m$, the share parameters satisfy the similar condition $\Gamma^2<0$ and therefore the components of the wave vector behave harmonically. Like the previous cases the exponential instability appears in the outer region.  In particular, in Fig. \ref{fig4} we present the plots of normalised total velocity perturbation and energy perturbation versus time for the same location of excitation $r=R/2$. The set of parameters is $\alpha=2.4$, $\beta=7.89$, $\gamma=-0.435$, $A = 0.3$ and $B=0.7$, $r_m=5\times 10^4$cm, $\upsilon_m = 8\times 10^3$cm s$^{-1}$, $\lambda_0=r_m/100$, $\upsilon_{x,y}(0) = 0.01$, $d(0) = 0$, $k_x(0) = 1$ and $k_y(0) = 0.1$. As it is clear from the graph, the hydrodynamic waves undergo strong instability and the wave pumps energy from the background flow, increasing its initial value by approximately $200$ times. 
\begin{figure}
\resizebox{0.5\textwidth}{!}{%
  \includegraphics{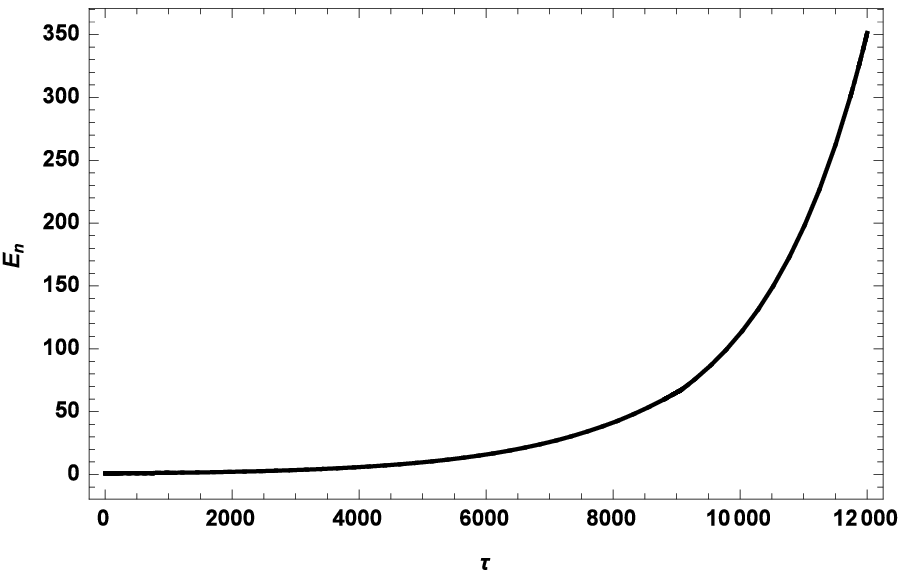}
  \includegraphics{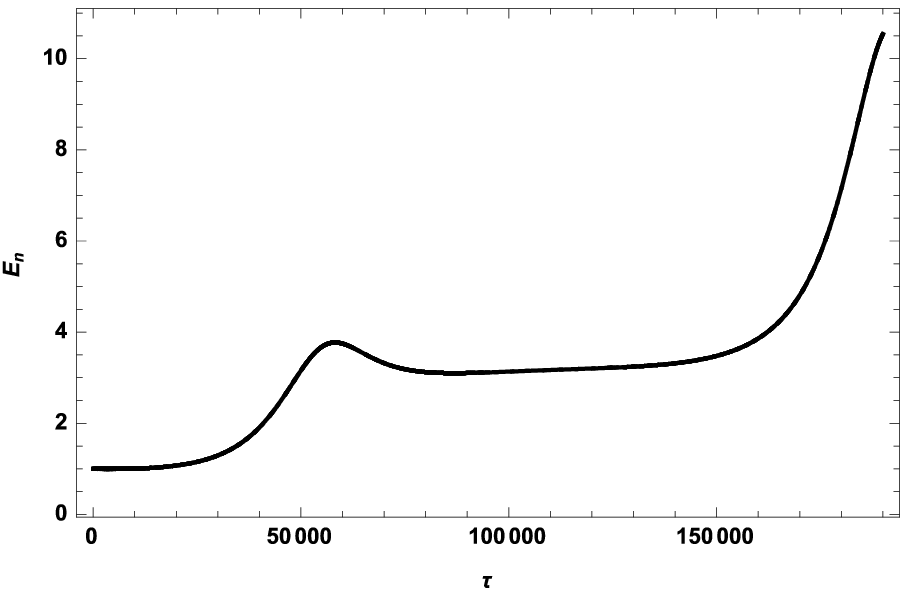}

}
\caption{For the modified Sullivan vortex we plot the temporal evolution of normalised values of energy perturbations in the outer region, $r = R/2$ (left panel) and the inner area, $r = R/3$, (right panel) of a tornado. The set of parameters is $\alpha=2.4$, $\beta=7.89$, $\gamma=-0.435$, $A = 0.3$ and $B=0.7$, $r_m=5\times 10^4$cm, $\upsilon_m = 8\times 10^3$cm s$^{-1}$, $\lambda_0=r_m/100$, $d(0) = 0.01$, $\upsilon_{x,y}(0) = 0$, $k_x(0) = 1$ and $k_y(0) = 0.1$. For the vertical components the parameters are following: $\frac{\partial U}{\partial r} = 0$, $k_z(0) = 0$ (left panel) and $\upsilon_z(0) = 0.00001$, $\frac{1}{K_x(0)C_s}\frac{\partial U}{\partial r} = -10^{-7}$, $k_z(0) = 0.1$ (right panel).}
\label{fig5}
\end{figure}

The Sullivan model has been modified in \cite{models} for a broader downdraft region occupying the tornado centre. The corresponding model parameters  are $\alpha=10$, $\beta=22$, $\gamma=-0.5$, $A = 0.091$ and $B=0.909$ (see Eq. (\ref{sullivan})). As it has been argued this model better describes the kinematic parameters of the so called "dust devil B" observed in Texas by means of a mobile W-band  \cite{devil}. 

In the context of the inner area, we have shown the situation drastically changes if a relatively small vertical component of velocity is added to the dynamical pattern. If this is the case the situation becomes three dimensional and the corresponding share matrix for the inner region is given by Eq. (\ref{S}). It is straightforward to check that if a small fraction of the vertical component is present the instability appears even inside the inner area of a tornado. 

In Fig. \ref{fig5} we plot the temporal evolution of normalised values of energy perturbation respectively in the outer region, $r = R/2$, (left panel) and the inner area, $r = R/3$, (right panel) of a tornado for the modified Sullivan vortex. In these examples initially only sound waves are excited. The set of parameters is $\alpha=2.4$, $\beta=7.89$, $\gamma=-0.435$, $A = 0.3$ and $B=0.7$, $r_m=5\times 10^4$cm, $\upsilon_m = 8\times 10^3$cm s$^{-1}$, $\lambda_0=r_m/100$, $d(0) = 0.01$, $\upsilon_{x,y}(0) = 0$, $k_x(0) = 1$ and $k_y(0) = 0.1$. For the vertical components the parameters are following: $\frac{\partial U}{\partial r} = 0$, $k_z(0) = 0$ (left panel) and $\upsilon_z(0) = 0.00001$, $\frac{1}{K_x(0)C_s}\frac{\partial U}{\partial r} = -10^{-7}$, $k_z(0) = 0.1$ (right panel). As it is clear from the plots, due to the velocity shear, initially excited sound mode becomes nonmodally unstable. As a result, for $\tau = 12000$ energy increases by the factor of $\sim 350$ in the outer region, and for $\tau = 180000$ - by the factor of $7$ in the inner region. The latter dimensionless time corresponds to approximately $10$ minutes, which might be interesting, because the average time of distraction of tornadoes is of the same order of magnitude.

 \section{Conclusion}
 
 By considering the typical kinematic parameters of tornadoes we have examined several vortex models to study how efficient the non-modal velocity SFs are. For this purpose we considered the Euler equation, continuity equation and the equation of state, linearised them and applied the mathematical model of SFs to tornadoes.
 
 As a first example we applied the approach to the Rankine vortex. It has been analytically shown that the wave vector components do not exhibit exponential growth in the inner area of a tornadoe. For the outer region, by perturbing only velocity components, we have shown that amplitudes of initially excited sound waves exponentially amplify.
 
 As a next step, we considered the Burgers-Rott model, which describes the one-celled tornadoes better than the Rankine vortex. It has been found that, as in the previous model, initial perturbation of velocity components leads to generation of non-modal SF instabilities. For example y exciting the sound waves the physical system undergoes instability. In particular, as it is evident from the plots energy of the waves increases by a factor of $400$ during $\tau = 12000$, which for the implied tornado parameters ($R = 500$m and $\upsilon_m=80$m/s) corresponds to approximately $0.5$min, meaning that the instability is very efficient.
 
 For the two-celled tornadoes the best kinematic fit can be achieved by the Sullivan and modified Sullivan vortex models. We have shown that even a negligible fraction of $\partial U_z/\partial r$ is enough to "switch" the instability in the mentioned area for the modified Sullivan vortex.
 
 All models exhibit the non-modal velocity SF instability. Generally speaking, it is evident that enhanced waves pump energy from the background flow and therefore, the mentioned instability might strongly influence the overall dynamics of tornadoes up to a moment of their destruction. On the other hand, in this paper we studied relatively simplified scenarios. In particular, we did not take into account the dissipation effects related to viscosity and the role of dust particles has not been examined as well. The present work is a first attempt to show importance of the velocity SF non modal instabilities in tornadic structures and this approach should be generalised by taking into account the aforementioned factors.

\section{Authors contributions}
All the authors were involved in the preparation of the manuscript.
All the authors have read and approved the final manuscript.

 \section*{Acknowledgments}
The research of MA was supported by the Knowledge Foundation at the Free University of Tbilisi and the research of ZO was supported by the Shota Rustaveli National Science Foundation grant (DI-2016-14). 

%
%

\end{document}